\documentclass[aps,prd,twocolumn,10pt,superscriptaddress,showkeys]{revtex4-2}

\usepackage{graphicx}
\usepackage{dcolumn}
\usepackage{bm}
\usepackage{amssymb}
\usepackage{epsfig} 
\usepackage{amsmath} 
\usepackage{graphicx}
\usepackage{color}
\usepackage{float}
\usepackage{xcolor}
\usepackage{mathtools}
\usepackage{amssymb}
\usepackage{enumitem}
\usepackage{natbib}
\usepackage[colorlinks=True, citecolor=blue,linkcolor=blue, urlcolor=blue]{hyperref}

\definecolor{red}{rgb}{0.8,0,0}
\definecolor{violet}{rgb}{0.4,0,0.4}
\definecolor{green}{rgb}{0,0.5,0.0}
\definecolor{navy}{rgb}{0.0,0.0,0.6}
\definecolor{orange}{rgb}{0.8,0.2,0.0}

\newcommand{\physrep}{Phys. Rep.}
\newcommand{\apjl}{Astro. Phys. J. Lett.}
\newcommand{\mnras}{Mon. Not. Roy. Astron. Soc.}

\newcommand{\physscr}{Phys. Scr.}

\newcommand{\apss}{Astrophys. Space Sci.}

\usepackage[normalem]{ulem}

\begin{document}

\preprint{APS/123-QED}

\title{Modeling large glitches with core superfluidity in a hybrid star}

\author{Yogeesh N}
\email{p23ph0018@iitj.ac.in}

\author{Anil Kumar}
\email{anil.1@iitj.ac.in}

\author{Monika Sinha}
\email{ms@iitj.ac.in}

\affiliation{Department of Physics, Indian Institute of Technology Jodhpur, Rajasthan, India}

\date{\today}

\begin{abstract}
Many pulsars exhibit a peculiar behavior in their pulse profile of a sudden increase in their rotational period, which is popularly known as a pulsar glitch. Some of them show giant glitches with relative amplitude $\Delta\Omega/\Omega \sim 10^{-6}-10^{-5}$. With the model of pinned neutron vortices inside the neutron star crust, this large glitch cannot be explained so far. However, the increasing evidence of massive pulsars indicates the appearance of exotic degrees of freedom in the inner core of the pulsars. Given this, we consider the pulsar as a hybrid star. This model opens up the possibility of vortex-pinning inside the core.  Under the Gibbs equilibrium conditions, it is possible for hadrons and the quark phase to coexist. Due to the global charge neutrality condition, quark pasta structures are formed in the background of hadronic matter. We consider these pasta structures as pinning sites of superfluid vortices. We show that considering the core contribution, our calculations come to be of the order of $\Delta\Omega/\Omega \sim 10^{-6}$, which is close to the observations shown by the Vela-like pulsars.
\end{abstract}

\maketitle

\section{Introduction \label{Introduction} }

Compact stars (CSs) are mostly observed as radio or x-ray pulsars. The pulsars are thought to be rotating CS with a high surface magnetic field ranging from $10^8 - 10^{16}$ G \cite{Thompson&Duncan1993,Price2006,Lattimer&Prakash2004}. They are very stable rotators with a small steady decrease in their rotational frequency. However, some of them exhibit a sudden increase in rotational frequency, which is popularly known as a pulsar glitch.  These are very intriguing and rare phenomena, have been observed in only $10\%$ of pulsars. Many studies have shown that glitches occur in relatively younger isolated pulsars of age between $10^3$ and $10^5$ years \cite{Yu_2012,Basu_2022,Liu_2025}. The observed fractional glitch size ranges from $\Delta\Omega/\Omega\sim10^{-9}-10^{-5}$ \cite{Espinoza_2011,Yu_2012}.

The origin of pulsar glitches is a long-standing question. Initially, two mechanisms were proposed to explain pulsar glitches. The first mechanism is to use star quakes in neutron stars, which builds up strain in the crust, resulting in a sudden rearrangement of the moment of inertia of the star \cite{Ruderman_1969,Baym_1971,Ruderman_1991}. However, subsequent glitch observations of the Vela pulsar proved that the starquake model could not satisfy their glitch frequency and size. The second mechanism is based on the sudden transfer of angular momentum from the crustal neutron superfluid to the rest of the star \cite{Anderson_1975,Ruderman_1976}. In this model, the glitch phenomenon is explained by considering the matter inside the CS to be in a superfluid state. Due to rotation, superfluid matter forms vortices that are pinned inside the crust. During the gradual spinning down of the pulsar, when these pinned vortices become unpinned, a significant amount of angular momentum is transferred to the CS, resulting in a glitch. In general, most of the works on this theory consider the pinning in the crust of the star \cite{Anderson_1975,Ruderman_1976,Pizzochero2011}. However, the angular momentum transfer from the crust part alone is not sufficient to explain the large glitches \cite{Andersson_2012,Chamel_2012} observed from many Vela-like pulsars \cite{Dodson_2007,Palfreyman_2018,Crab2018,J0357}.

Recent astrophysical observation on pulsars reveals the existence of massive pulsars, which indicate the possibility of the appearance of exotic degrees of freedom inside the interior of the CS \cite{Lattimer&Prakash2016,Baym2018,Annala2020}. This opens up the possibility of the existence of hybrid stars (HS) composed of deconfined strange quark matter (SQM) at the center of the star and surrounded by nucleonic matter up to the crust \cite{Anil2023}. It is believed that the conditions for the transition from hadronic matter (HM) to deconfined SQM are met in the interior of a CS where matter is at several times of nuclear saturation density $\rho_0$ \cite{Glendenning}. Considering the first-order phase transition from HM to deconfined SQM with Gibbs' criterion, a region within the star occurs where the two phases coexist, which is called a mixed phase \cite{Glendenning1992}. For the energy minimization, this phase occurs to be inhomogeneous with quark phase structure immersed in hadronic phase analogous to the nuclear pasta phase found in the inner crust of neutron stars \cite{Lattimer&Prakash2004}. In this scenario, in addition to the crust, this mixed-phase region within the star can act as a pinning region for the superfluid vortices. We consider the contribution of unpinning of the vortices from this region in the core of the star to explain large glitches. 

Naturally, in our model of glitch, the star is supposed to be HS composed of SQM at the core of the star surrounded by the nuclear matter (NM) composed of nucleons and leptons. The stellar model employed in this work is discussed in Sec.~\ref{sec:star_model}. In Sec.~\ref{sec:glitch model}, we briefly describe the glitch model incorporating core superfluidity. Finally, our results and conclusions are presented in Secs.~\ref{sec:results} and \ref{sec:conclusion}, respectively.

\section{Star model}\label{sec:star_model}
\subsection{Matter}\label{subsec:matter}

We consider that at lower density, matter is composed of nucleons along with some leptons. For the equation of state (EOS) of NM, we use the covariant density functional (CDF) model \cite{RING1996193,Glendenning:1997wn} with DDME2 parametrization \cite{DDME2}. The model assumes interaction between baryons is mediated via isoscalar-scalar $\sigma$, isoscalar-vector $\omega$ and isovector-vector $\rho$ mesons. We consider the matter NM to be composed of neutrons, protons, and electrons. The Lagrangian density of NM with only nucleons {within the CDF model is
\begin{align}
    {\cal L}_{NM} &= \sum_{n,p} \bar{\psi}_N(i\gamma_{\mu} D^{\mu} - m^{*}_N) \psi_N - \frac{1}{2}\partial^{\mu}\sigma\partial_{\mu}\sigma - \frac{1}{2}{m_{\sigma}^2\sigma^2} \nonumber  \\
    & - \frac{1}{4}\omega^{\mu\nu}\omega_{\mu\nu} + \frac{1}{2}m_{\omega}^2\omega^\mu\omega_\mu - \frac{1}{4}\boldsymbol{\rho^{\mu\nu}}\cdot\boldsymbol{\rho_{\mu\nu}} \nonumber  \\
    & + \frac{1}{2}m_{\rho}^2\boldsymbol{\rho^\mu}\cdot\boldsymbol{\rho_\mu} + \bar{\psi}_e\left(\gamma_{\mu}i\partial^\mu -m_e\right)\psi_e,
\end{align}
Here, $\psi_N$ and $\psi_l$ are the fields of nucleons and leptons, with their masses $m_n$ and $m_l$, respectively. The covariant derivative is $D_\mu = \partial_\mu + ig_{\omega N} \omega_\mu + ig_{\rho N} \boldsymbol{\tau}_{N3} \cdot \boldsymbol{\rho}_{\mu}$ and the effective mass of nucleon is $m_{N}^* = m_N - g_{\sigma N}\sigma$. Here, the vector fields are $\omega_{\mu \nu}  = \partial_{\mu}\omega_{\nu} - \partial_{\nu}\omega_{\mu}$ and $\boldsymbol{\rho}_{\mu \nu}  = \partial_{\nu}
\boldsymbol{\rho}_{\mu} - \partial_{\mu}\boldsymbol{\rho}_{\nu}$. The coupling parameter of nucleon with the $i$th meson is density-dependent as \citep{TYPEL1999331,PhysRevC.66.024306}
\begin{equation}
g_{i N}(n)= g_{i N}(n_{0}) f_i(x) \quad \quad \text{for }i=\sigma,\omega
\end{equation}
where the function is given by
\begin{equation}\label{eqn.func}
f_i(x)= a_i \frac{1+b_i (x+d_i)^2}{1+c_i (x +d_i)^2}
\end{equation}
where $x=n/n_0$ and $a_i$, $b_i$, $c_i$, $d_i$ are constants, describing properties at saturation density ($n_0$).
But, for the $\rho$-meson, the density-dependent coupling parameter is 
\begin{equation}
g_{\rho N}(n)= g_{\rho N}(n_{0}) e^{-a_{\rho}(x-1)}
\end{equation}
The chemical potential of a nucleon can be obtained as
\begin{equation}
\mu_N = \sqrt{{p_{f_N}}^2 + {m_N^*}^2} + g_{\omega N}\omega_0 + g_{\rho N} \boldsymbol{\tau}_{N3}{\rho}_{03} + \Sigma^r
\end{equation}
where $\Sigma^r$ is the rearrangement term that arises due to the density-dependent parametrization given by \citep{PhysRevC.64.025804}
\begin{equation}
\begin{aligned}
\Sigma^{r} & = \sum_{N} \left[ \frac{\partial g_{\omega N}}{\partial n}\omega_{0}n_{N} - \frac{\partial g_{\sigma N}}{\partial n} \sigma n_{N}^s + \frac{\partial g_{\rho N}}{\partial n} \rho_{03} \boldsymbol{\tau}_{N3} n_{N} \right]
\end{aligned}
\end{equation}
In this rearrangement, term $n$ denotes the number density and $n^s$  denotes the scalar number density. The energy density of NM can be calculated as
\begin{align}
\varepsilon_{\mathrm{NM}} &= \frac{1}{\pi^2} \bigg(\sum_N \int_{0}^{p_{f_N}}k^2\sqrt{k^2+{m_N^*}^2}dk \nonumber\\ 
&+\sum_l \int_{0}^{p_{f_l}}k^2\sqrt{k^2+{m_l}^2}dk \bigg) + \frac{1}{2}m_{\sigma}^2 \sigma^{2} \nonumber\\ &+ \frac{1}{2} m_{\omega}^2 \omega_{0}^2 + \frac{1}{2}m_{\rho}^2 \rho_{03}^2 
\end{align}
Here, $p_f$ is the Fermi momentum, and the pressure can be evaluated by the relation 
\begin{equation} 
 p_{\mathrm{NM}} = \sum_{i=N,l} \mu_in_i - \varepsilon_{\mathrm{NM}}.
\end{equation}}

We consider the coupling constants and masses of the mesons from the DDME2 parametrization \cite{PhysRevC.71.024312} as given in Table \ref{tab:DDME2}.
\begin{table}[ht]
\centering
\begin{tabular}{cccccccc|}
\hline
\hline
$i$ &$m_i$ & $g_i(\rho_{\text{sat}})$ & $a_i$ & $b_i$ & $c_i$ & $d_i$ \\
\hline
$\sigma$ & 550.1238 & 10.5396 & 1.3881 & 1.0943 & 1.7057 & 0.4421\\
$\omega$ & 783.0000 & 13.0189 & 1.3892 & 0.9240 & 1.4620 & 0.4775\\
$\rho$ & 763.0000 &  3.6836 & 0.5647 & & &\\
\hline
\hline
\end{tabular}
\caption{DDME2 parametrization. $\rho_{sat}$ is baryonic saturation density and $m_i$ is the mass of meson $i$.}
\label{tab:DDME2}
\end{table}

With the increase in density, NM may get deconfined, and the matter goes under a phase transition to SQM. We consider that the transition happens via a first-order phase transition with Gibbs construction \cite{Glendenning}. During this phase transition, both phases coexist for a certain range of density, which gives rise to a mixed phase \cite{Glendenning:1997wn}.  Here, we do not include heavier baryons at high densities, as their presence is suppressed by the appearance of quarks in the charge-neutral matter \cite{PhysRevD.76.123015}. In this mixed-phase region, the charge neutrality condition becomes as
\begin{equation}
    \chi\rho_c^{SQM} + (1-\chi)\rho_c^{NM} = 0.
\end{equation}
Here, $\rho_c$ is charge density and $\chi$ is the fraction of SQM by volume as
\begin{align}
    \chi = \frac{V_{SQM}}{V_{SQM} + V_{NM}} \nonumber.
\end{align}
The baryonic density of matter is
\begin{equation}
    \rho_{MP} = \chi\rho_{SQM} + (1-\chi)\rho_{NM},
\end{equation}
where, $\rho_{SQM}$ is $(n_u+n_d+n_s)/3$, and $n$ is the number density of quarks.

\begin{figure}
    \centering
    \includegraphics[width=0.85\linewidth]{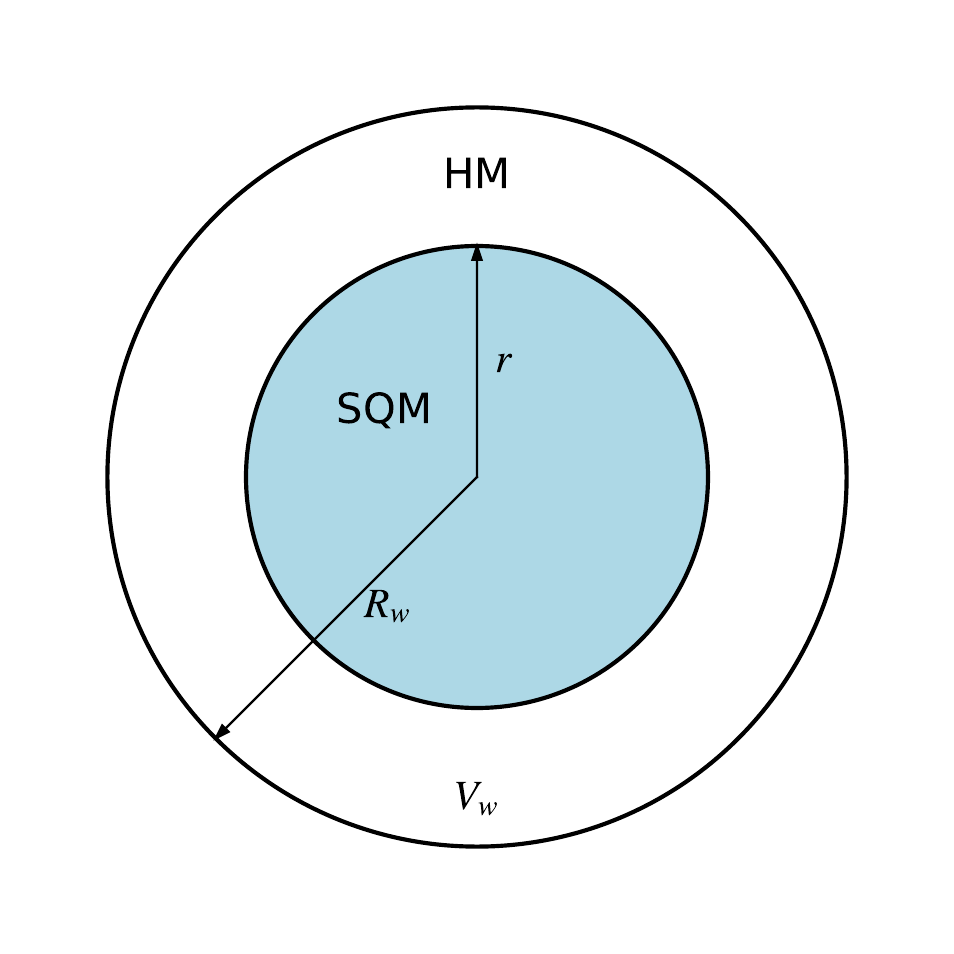}
    \caption{Within the mixed-phase region, we consider a structured configuration where spherical droplets of SQM, each with radius \(r\), are embedded in a surrounding medium of NM. The two components are separated by well-defined, sharp boundaries \cite{Maruyama2006}.}
    \label{fig:sharp bondary}
\end{figure}

As already mentioned, this mixed state is inhomogeneous. Due to the requirement of energy minimization, the mixed phase of confined and deconfined matter is expected to form a crystalline lattice \cite{Glendenning1992}. The formation of a lattice in the mixed phase arises due to the influence of the isospin symmetry energy in nuclear matter. At lattice points, the rarer phase with several geometric structures is immersed in the dominant phase. With varying densities, different structures emerge, forming a phase analogous to the nuclear pasta phase found in the inner crust of neutron stars. A range of complex structures exists and are typically referred to as slab, rod (or tube), and droplet (or bubble),  corresponding to one, two, and three dimensions, respectively \cite{Avancinin_2008}. In the energy minimization method, we describe mixed phase within the Wigner-Seitz (WS) approximation, where the whole space is divided into equivalent cells with a geometric symmetry \cite{Ju_2021}. We consider that the hadronic and quark phases are spatially separated within the WS cell with volume $V_W$: a quark-phase lump is embedded in the hadronic background, or vice versa, as shown in Fig. \ref{fig:sharp bondary}. The radius of the  WS cell is 
\begin{equation}
    R_{WS} = \left(\frac{3}{4\pi n_b}\right)^{1/3},
\end{equation}
with its volume $V_W$ being equal to the inverse of the baryonic density $n_b$ \cite{Than_2010}.

The two phases are separated by a sharp boundary with surface tension
{$\sigma = \Sigma_{i=u,d,s} \sigma_i$ where the contribution to the surface tension for each quark species $i$ is calculated as \cite{Xia_2020}
\begin{equation}
\begin{split}
\sigma_i = \frac{g_i m^3_i}{48\pi^2} \Big[ & (4v_{f_i} - 3\pi)\sqrt{v^2_{f_i}+1} + 2\pi 
- 2sinh^{-1}(v_{f_i}) \\
& + 2(v^2_{f_i}+1)^{3/2} cot^{-1}(v_{f_i}) \Big].
\end{split}
\label{eq:surface tension}
\end{equation}
Here $v_{fi} = p_{fi}/m_i$ with $p_{fi}$ being the Fermi momentum of quarks (\textit{i=u,d,s}). The total surface tension \textbf{$\sigma$} is the sum of all three quark contributions. The surface tension allows us to determine the structure of the hadron-quark mixed phase.

The total energy of mixed phase is \cite{Ravenhall1983,Xia_2020}, 
\begin{equation}
\varepsilon_{MP}=\varepsilon_{\rm s}+\varepsilon_{\rm C}+\chi \varepsilon_{SQM}+(1-\chi)\varepsilon_{NM},
\end{equation}
where surface energy per unit volume
\begin{equation}
\varepsilon_{\rm s}=d\chi\frac{\sigma}{r},
\end{equation}
and Coulomb energy per unit volume
\begin{equation}
\varepsilon_{\rm C}=\frac{2\pi\alpha\chi r^{2}(\rho_{c}^{SQM})^2}{(1-\chi)^{2}(d+2)}\left[\frac{2}{d-2}\left(1-\frac{d}{2}\chi^{1-\frac{2}{d}}\right)+\chi\right],
\end{equation}

for $d=2$ \footnote{With electronic communication with authors of \cite{Xia_2020}}\cite{Mariani_2024},
\begin{equation}
 \varepsilon_{\rm C}=\frac{2\pi\alpha\chi r^{2}(\rho_{c}^{SQM})^2}{(1-\chi)^{2}(d+2)}\left[\chi-1-ln(\chi)\right].
\end{equation}
Here $d$ is the dimensionality and $r$ is the radius of the quark structures.

By taking the derivatives of the mixed-phase energy density $\varepsilon_{MP}$ with respect to the independent variables ($r,d,\chi,n_q,\rho_{c}^{SQM}$), we obtain a set of conditions that define the conditions for mixed phase with structures. The equilibrium conditions are \cite{Xia_2020}
\begin{equation}
\varepsilon_s = 2\varepsilon_C,
\end{equation}

\begin{equation}
P^{\text{SQM}} - P^{\text{NM}} = \frac{d \Sigma \left( d\chi^{\frac{2}{d}} - d + \chi^{\frac{2}{d}-1} - \chi^{\frac{2}{d}} + 1 - \chi \right)}
{r (1 - \chi)\left( d\chi^{\frac{2}{d}} - d - 2\chi^{\frac{2}{d}} + 2\chi^{\frac{2}{d}-1} \right)}
,
\end{equation}

for $d=2$,
\begin{equation}
P^{\text{SQM}} - P^{\text{NM}} = \frac{d \Sigma \left(\chi^2-3\right)}
{2r (1 - \chi)},
\end{equation}

\begin{equation}
\frac{(d^3 - 12d + 16)\chi - 16}{(2d^2 - 8)\ln(\chi) + d^3 - 12d}\chi^{\frac{2}{d}-1} = 1,
\end{equation}

\begin{equation}
\mu_B^{\text{SQM}} = \mu_B^{\text{NM}},
\end{equation}
\begin{equation}
\mu_e^{\text{SQM}} - \mu_e^{\text{NM}} = \frac{d\sigma}{r\rho_{c}^{SQM}}.
\end{equation}

For the SQM, we opt to choose the vector MIT bag (vBAG) model \cite{2015ApJ...810..134K,2021PhyS...96f5303L} because it goes well with recent astrophysical observations \cite{Anil2023}. It includes vector interactions analogous to $\omega$ meson in NM. The Lagrangian density of SQM is
\begin{equation}
\begin{aligned}
{\cal L}_{\mathrm{SQM}} = & \sum_{q = u,d,s} \Big[ 
\bar{\psi}_q \big\{ \gamma_{\mu}(i\partial^\mu - g_{qV}V_\mu) - m_q \big\} \psi_q 
- B \Big] \\
& \times \Theta(\bar{\psi}_q \psi_q) 
- \frac{1}{4}(\partial_{\mu}V_{\nu} - \partial_{\nu}V_{\mu})^2 \\
& + \frac{1}{2} m_V^2 V_\mu V^\mu 
+ \bar{\psi}_e \left( \gamma_{\mu} i\partial^\mu - m_e \right) \psi_e
\end{aligned}
\end{equation}
where $\Theta$ is the Heaviside step function, which vanishes outside the bag and remains unity inside the conventional bag, and $g_{qV}$ is the coupling parameter with the quark. The difference between the pressure inside this bag and the true vacuum is the bag parameter B. With this model, the energy density of the quark matter can be calculated as 
\begin{equation}
\begin{aligned}
\varepsilon_{\mathrm{SQM}} = & \; \frac{3}{\pi^2} \sum_q \int_{0}^{k_{f_q}} 
\left( \sqrt{k^2 + m_q^2} + m_V V_0 \right. \\
& \left. \times \sqrt{G_V} \right) k^2 \, dk 
- \frac{1}{2} \left( m_V V_0 \right)^2 + B,
\end{aligned}
\end{equation}

The parameter $G_V$ is defined as $(g_{qV}/m_V)^2$, where $m_V$ is the mass of the vector meson. The term $m_VV_0$ is 
\begin{equation}
\begin{aligned}
m_VV_0 = \sqrt{G_V}\left(\sum_qn_q\right)
\end{aligned}
\end{equation} 
and pressure as:
\begin{equation}
\begin{aligned}
p_{\mathrm{SQM}}= \sum_{q}\mu_qn_q - \varepsilon,
\end{aligned}
\end{equation} 
where $n_q$ denotes number density of quark $q$. For the details, the reader may look into the Ref. \cite{2022MNRAS.513.3788K}.

The equation of state (EOS) of matter is shown in the Fig. \ref{fig:EoS}. At lower densities, the matter is composed of only nucleons, and depending on the parameter space of the vBAG model described above, the deconfined SQM appears after some density $\rho_{MP}$, initiating the mixed phase. The variation of pressure $P$ with energy density $\epsilon$ for two values of $G_V$(0.2 fm$^2$ and 0.3 fm$^2$ ) and two values of $B^{1/4}$(155 MeV and 170 MeV)are shown in Fig. \ref{fig:EoS}.  
\begin{figure}[h]
    \centering
    \includegraphics[width=\linewidth]{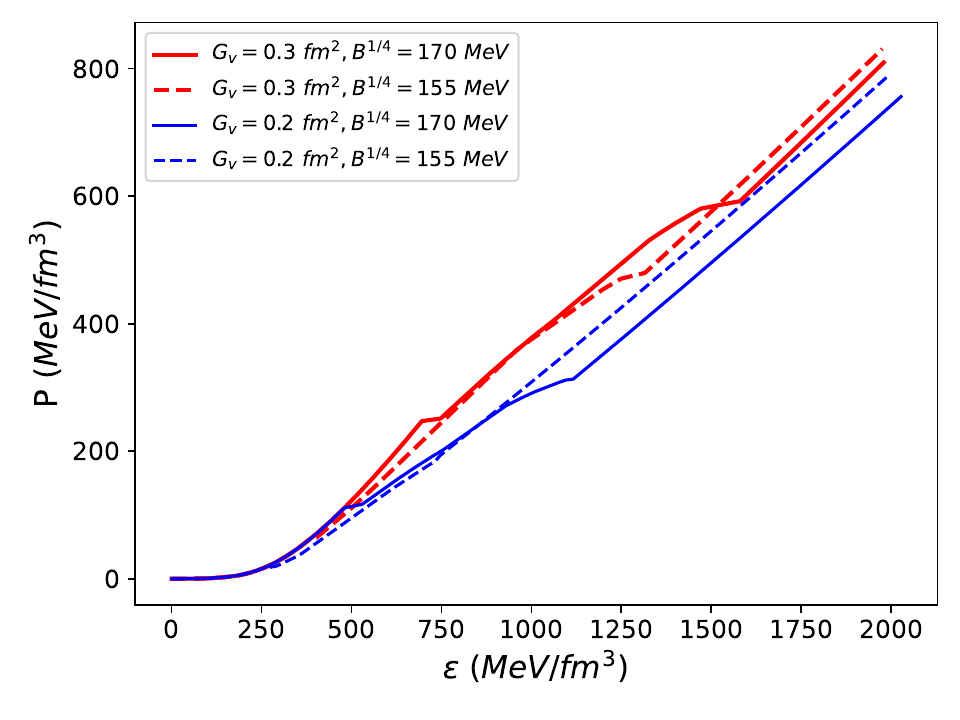}
    \caption{Variation of pressure with energy density.}
    \label{fig:EoS}
\end{figure}

As discussed, in the mixed phase, we adopt the concept of continuous dimensionality of mixed-phase structures as proposed by \cite{Ravenhall1983} and later employed in \cite{Xia_2020}, allowing for a smooth transition between idealized geometries such as droplets, rods, and slabs. The intermediate values of dimensionality in the range d=3 to d=1 correspond to evolving structures that lie between idealized drops, rods, and slabs, respectively, reflecting a continuous transformation in the geometry of the mixed phase. 

In our case the mixed phase begins to appear at density  $n_B \approx 0.495~\mathrm{fm}^{-3}$ with droplike structures, the rodlike structures appear at the density  $n_B \approx 0.552~\mathrm{fm}^{-3}$ and slablike structures begins at  $n_B \approx 0.656~\mathrm{fm}^{-3}$ whereas tube and bubble like structures appear at $\approx 0.845~\mathrm{fm}^{-3}$ and $\approx 0.878~\mathrm{fm}^{-3}$, respectively. Similar kind of trend can be found in Table 1 of \cite{2021ApJ...923..250J}. The variation of dimensionality of these structures in the mixed phase region is shown in Fig. \ref{fig:MP-Params} for the EoS  $G_v=0.2$ fm$^2$ and $B^{1/4}=170$ MeV.

\begin{figure}[H]
    \centering
    \includegraphics[width=\linewidth]{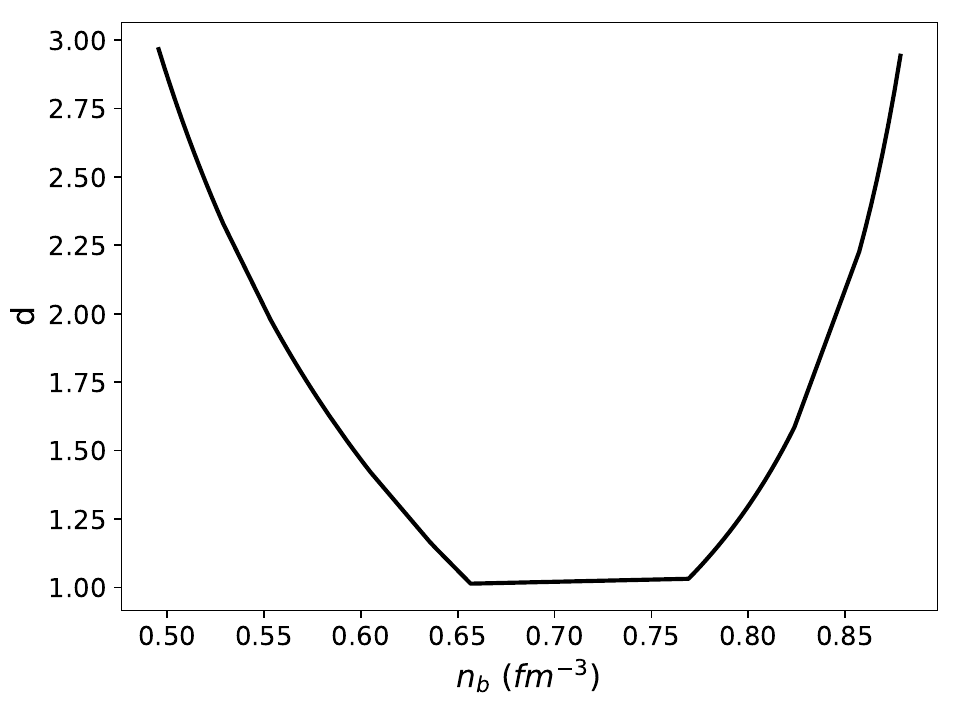}
    \caption{Variation of the dimensionality (d) with respect to baryon number density. }
    \label{fig:MP-Params}
\end{figure}

\subsection{Stellar structure}\label{sec: Stellar structure}
 With the matter as discussed in the Sec. \ref{subsec:matter}, the star core can be divided into three parts; 1) the outer core, composed of pure NM, 2) the inner core, composed of SQM in mixed phase, and 3) the innermost core, composed of pure SQM. In the inner core, when the matter density is several times $\rho_0$, mixed phase with NM and SQM appears with the matter composition outlined in the previous section.  Immediately after the outer core, the hadronic and quark phase coexist, and if the star is sufficiently massive, in the innermost core pure quark phase appears \cite{Glendenning1992,Anil2023}. Consequently, the interior of the star in our model is divided into several regions depending on the mass of the star, as shown in a schematic diagram by Fig. \ref{fig:stellarstucture}. 
\begin{figure}[h]
    \centering
    \includegraphics[width=0.95\linewidth]{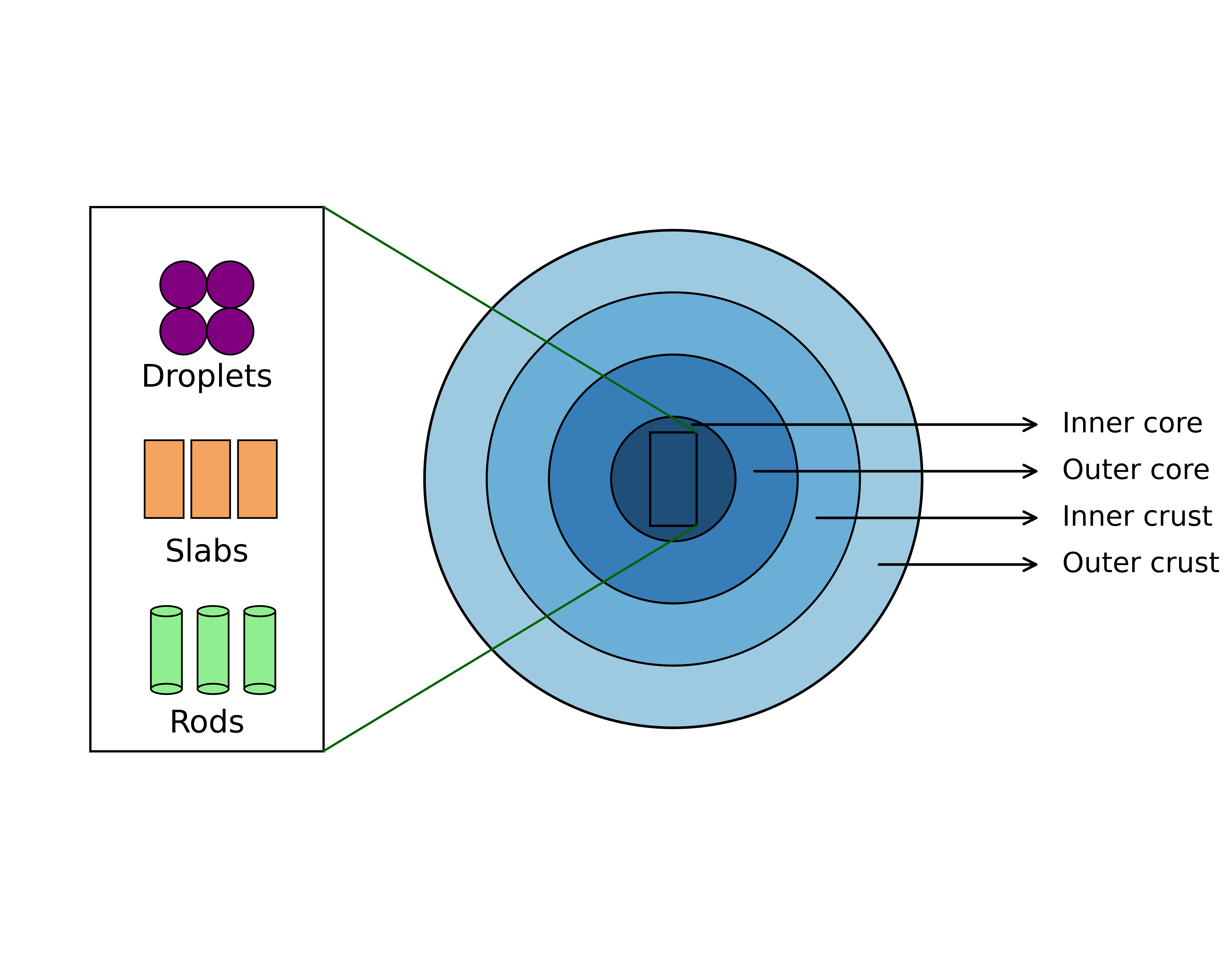}
    \caption{Interior of a hybrid star. The quarks structures which are presumed to be inside the core are depicted as well.}
    \label{fig:stellarstucture}
\end{figure}

With EOSs shown in Fig. \ref{fig:EoS} we construct the mass-radius (M-R) relation of the HS, shown in Fig. \ref{fig:MR}.

\begin{figure}[h]
    \centering
    \includegraphics[width=1.0\linewidth]{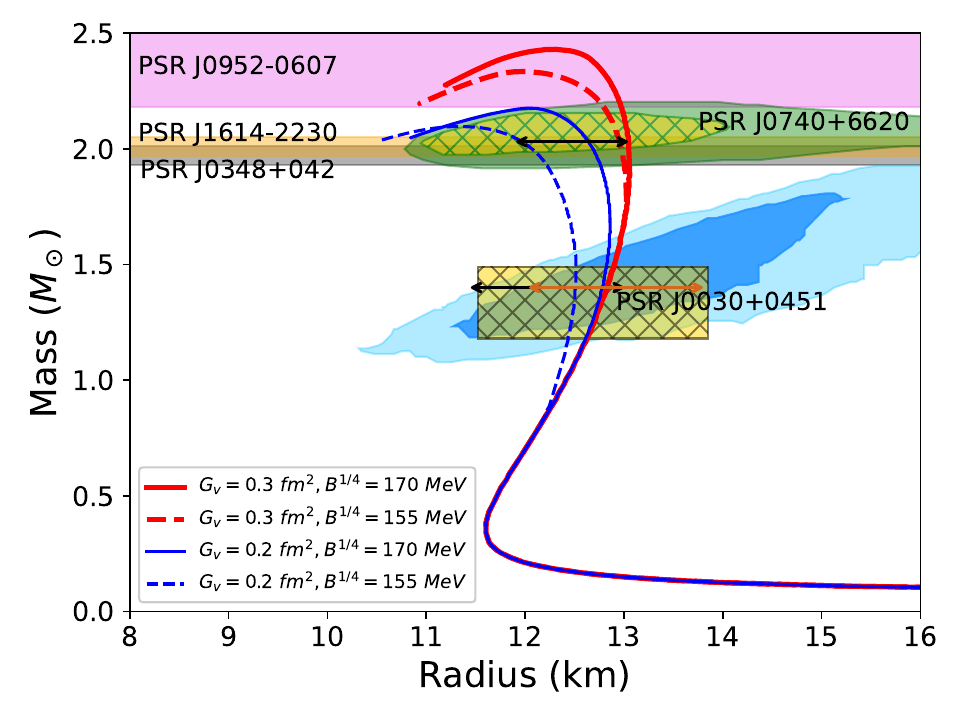}
    \caption{Mass-radius relation of mixed phase EOSs along with the mass and radius constraints.The contours represent constraints for PSR J0740+6620 \cite{Riley:2019yda, Riley:2021pdl} and PSR J0030+451 \cite{Miller:2019cac,Miller:2021qha}. The horizontal lines indicate radius constraints. The shaded region at the top corresponds to PSR J0952–0607 \cite{Romani:2022jhd}. }
    \label{fig:MR}
\end{figure}

\section{Glitch Model}\label{sec:glitch model}
We consider neutrons inside the star to be in the superfluid state \cite{Takatsuka_1971,Anderson_1975,Andersson_2012,Sedrakian:2018ydt}. Superfluids rotate by forming an array of vortices. A neutron vortex inside the star cannot be considered straight throughout the star with a realistic approach \cite{seveso2016mesoscopic}. In the inner core, due to the presence of structure in mixed phase, the straight part of the vortices experiences strong pinning \cite{Hirasawa_2002} as surface tension is strong enough for neutron superfluid vortices to be pinned to these pasta structures. The straight segments of a vortex line intersect nearly all nuclei along the major axis of the lattice, whereas the kinked segments are located primarily in the interstitial regions between lattice planes. As a result, the vortex is strongly pinned to the nuclei along its straight portions, but remains weakly or not pinned in the kinked regions. The length $H$ of the vortex up to which it can be considered straight is given by \cite{seveso2016mesoscopic}

\begin{equation}
    H = \frac{2TR^2_{WS}}{ |E_p|},
\end{equation}
where 
\begin{equation}
   T = \rho_n \frac{\kappa^2}{4\pi}\ln\left(\frac{l_v}{\xi}\right),
\end{equation}
is the tension on the vortex with $l_v$ the inter-vortex spacing and $E_P$ the pinning energy given by 
\begin{equation}
    E_p = f_p R_{WS} \xi .
\end{equation}

For any pulsar rotating with angular velocity $\Omega$, $l_v$ can be calculated as \cite{van_Eysden_2018}{
\begin{equation}
    l_v = 4 \times 10^{-3}\left(\frac{\Omega}{20\pi}\right)^{-\frac{1}{2}}cm.
\end{equation}}

Inside the inner core, mixed phases of NM and SQM appear with structures. In this phase in a WS cell, the hadronic and quark phases are assumed to be separated by a sharp interface characterized by a finite surface tension. A well-defined boundary is imposed between the two phases, and the associated surface energy is incorporated through a surface tension parameter $\sigma$. We consider that this surface tension is strong enough for neutron superfluid vortices to be pinned to these pasta structures with a pinning force per length \cite{Yasutake_2014}, 
\begin{equation}
    f_P = \sigma.
\end{equation} 

\begin{figure}[h]
    \centering
    \includegraphics[width=1.0\linewidth]{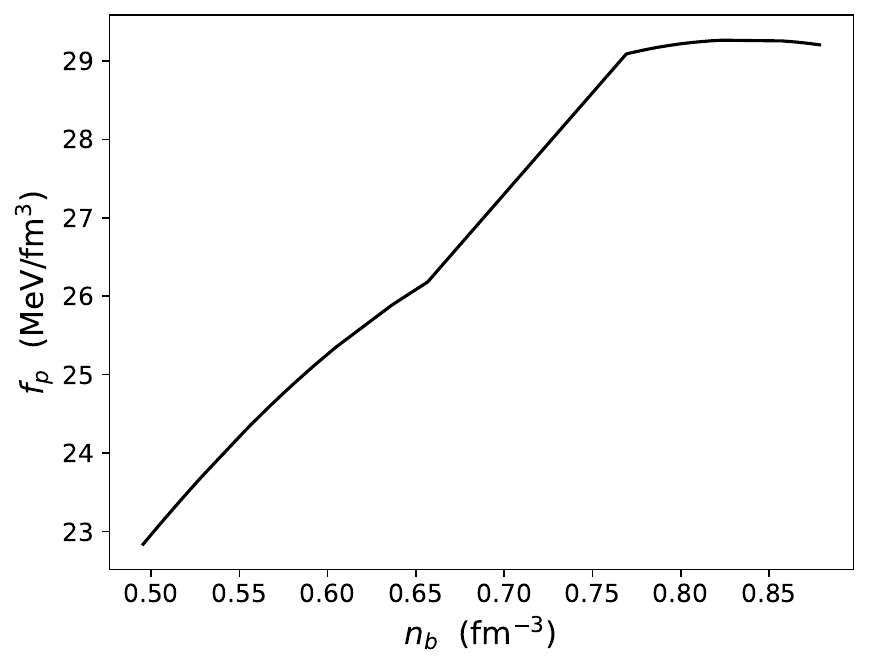}
    \caption{The variation of surface tension with the baryon density. }
    \label{fig:fp}
\end{figure} 

The variation of surface tension with baryon density is shown in Fig. \ref{fig:fp}

The pulsars gradually slow down with decreasing rotational speed. The vortex aerial density is directly related to the rotational frequency. So, it also tends to decrease. However, due to pinning, it can not decrease while keeping the rotational frequency of the superfluid surrounding the vortices intact \cite{Pizzochero2011}. Due to the velocity difference between the neutron superfluid and the vortex lines, the vortices experience a Magnus force per unit length, given as \cite{seveso2012},
\begin{equation}
    \textbf{f}_M =  \rho_n \, \mathbf{\kappa} \times (\textbf{v}_n - \textbf{v}_v),
\end{equation}
where $\mathbf{\kappa}$ is the quantum of circulation, \( \rho_n \) is the density of the neutron superfluid, and $\textbf{v}_n$ and $\textbf{v}_v$ are the velocities of the neutron fluid and the vortex line, respectively. When the velocity lag between the superfluid neutrons and the vortex line reaches a critical threshold \( \omega_{\text{cr}} \), the Magnus force acting on the vortex becomes equal in magnitude to the local pinning force. For a straight segment of vortex of length $H$ it can be expressed as
\begin{equation}
   \int^H_0 f_P dz = \int^H_0 f_M dz = \int^H_0 \kappa\rho_n x \omega_{cr} dz
\end{equation} 
where $x$ is the distance of the vortex from the rotational axis. This condition leads to vortex unpinning. The critical angular velocity lag for unpinning is therefore given by 
\begin{equation}
    \omega_{cr} =  \frac{\int^H_0 f_P dz}{\kappa x \int^H_0 \rho_n dz}
\end{equation}
  When vortices are unpinned, the arial density of vortices decreases, thereby transferring the stored angular momentum inside them to the star crust. Thus, a glitch of a sudden increase in star rotation occurs.

\subsection{Spin-up during the glitch}
The total angular momentum transferred from the pinned portion of the vortices to the star at the time of the glitch is given by
\begin{equation}
    \Delta L = \int^{R_{hq}}_{R_q} dI_n(x) \omega_{cr}(x),
\end{equation}
where the mixed phase region of the star is bound between the radii $R_q$ and $R_{hq}$, $dI_n(x)$ is the superfluid neutron moment of inertia responsible for the glitch at distance $x$ from the star rotation axis, and is given by
\begin{equation}
    dI_n(x) = \rho_n x^2 dx  x d\phi \int_0^Hdz = 2\pi x^3dx\int_0^H\rho_ndz.
\end{equation}
 Hence, the total angular momentum is \cite{Antonelli_2018} 
\begin{equation}
\begin{split}
    \Delta L = \frac{2\pi}{{\kappa } } \int^{R_{hq}}_{R_q}  dx x^2 \int^{H}_{0} f_p dz.
    \end{split}
\end{equation}

If the star's moment of inertia is $I$, the change in rotational speed of the star $\Omega_{gl}$ is 
\begin{equation}
    \Omega_{gl} = \frac{\Delta L}{I}.
\end{equation}

\subsection{Time interval between two subsequent glitches}
The glitch will occur after the rotational speed of the star crosses the  $\omega_{cr}$ and $\omega_{cr}$ varies with the radius. So, time required for the next glitch to build up is \cite{seveso2016mesoscopic}
\begin{equation}
    t_{gl} = \frac{(\omega_{cr})_{max}}{\dot{\Omega}_c},
\end{equation}where $\dot{\Omega}_c$ is the spin-down rate of the pulsar \cite{Pizzochero2011,seveso2012}.
This corresponds to the region of strongest pinning, which sets the threshold for large-scale vortex unpinning \cite{seveso2016mesoscopic}. Therefore, the time interval between glitches is  calculated using $(\omega_{cr})_{max}$.

\section{Results} \label{sec:results}

 From the Fig. \ref{fig:MR} it is evident that all the observed mass-radius constraints are satisfied by the M-R curve with $G_v=0.2$ fm$^2$ and $B^{1/4}=170$ MeV. For the rest of the calculation, we consider this parametrization.

As the matter inside the star is in a superfluid state, the neutrons rotate with the star, forming vortices as described above. These vortices will be pinned in the mixed phase region due to the surface tension associated with the structure in the mixed phase region. The surface tension increases with density, resulting in a significant rise in the vortex pinning energy. This enhanced pinning strength plays a crucial role in the glitch mechanism by anchoring superfluid vortices more effectively in the high-density regions of the star. Moreover, the contribution of surface tension from hadronic structures is neglected, as its magnitude is comparatively small \cite{Mariani:2023kdu}.

 Due to vortex pinning in the neutron star core, the superfluid vortices cannot immediately follow the spin-down of the star. This leads to the build-up of a rotational lag between the stellar core and the superfluid vortices. The maximum sustainable lag before vortices unpin is referred to as the critical lag $\omega_{cr}$. We calculate the $\omega_{cr}$ for which the Magnus force will be equal to the pinning force leading to unpinning of the vortices. The variation of $\omega_{cr}$ with the radius of the star of mass $1.4~M_\odot$ with the above-mentioned EOS parametrizations is shown in the fig \ref{fig:Critical lag}.
\begin{figure}[H]
    \centering
    \includegraphics[width=1.0\linewidth]{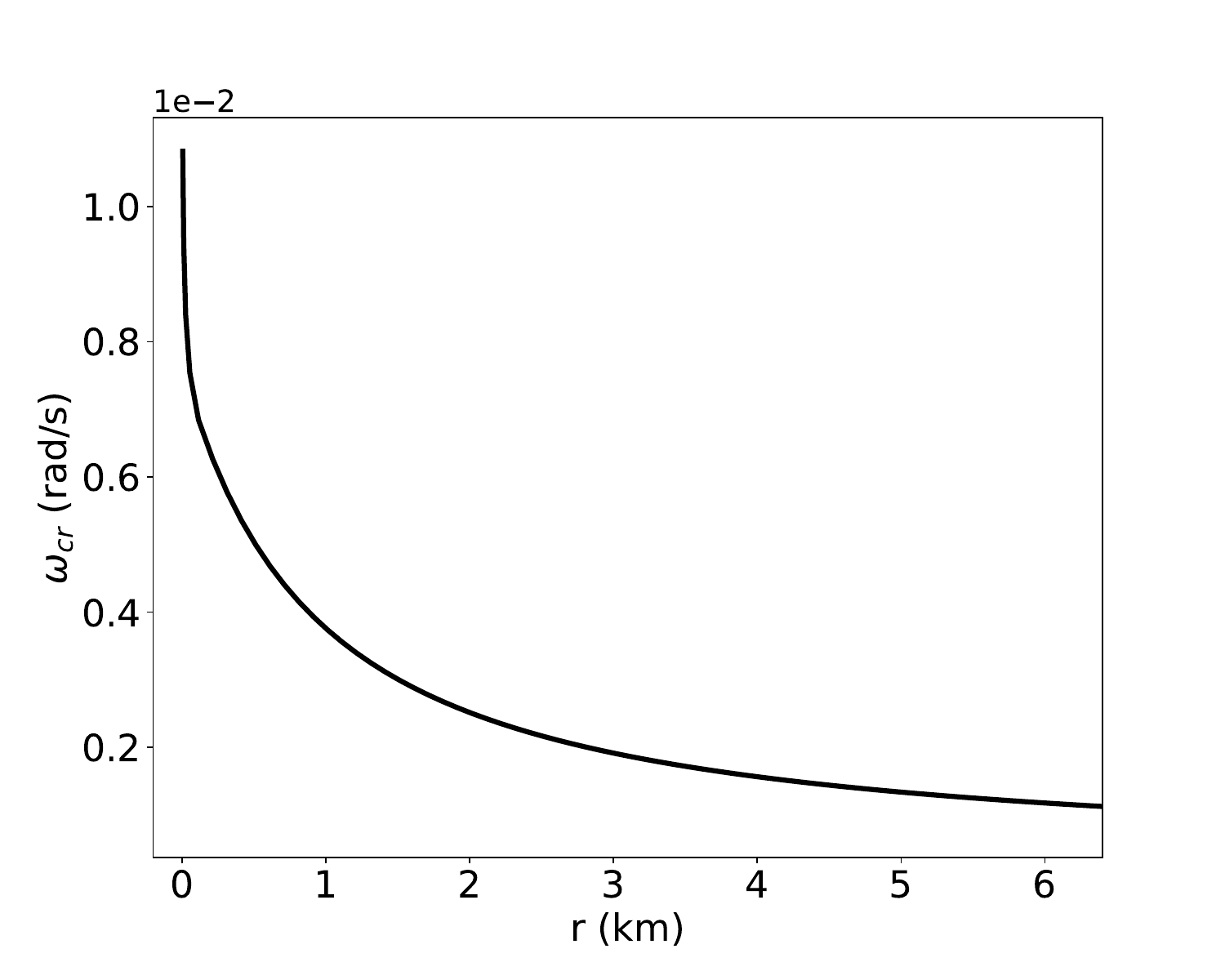}
    \caption{Variation of critical lag $\omega_{cr}$ with the radius of the star. }
    \label{fig:Critical lag}
\end{figure} 

\vspace{0.5cm}
\begin{table}[htbp]
\centering
\renewcommand{\arraystretch}{1.2}
\begin{tabular}{ccccc}
\hline\hline
Mass & $\Delta$ (MeV) & $\Delta \mathrm{L}$ (ergs s) & $\Delta \Omega_{gl}$ (rad/s) & $\Delta \Omega_{gl}/\Omega$ \\
\hline
{1.4 $M_\odot$} & 0.1 & $2.475\times10^{41}$ & $1.306\times10^{-4}$ & $1.856\times10^{-6}$ \\
 & 0.3 & $7.426\times10^{41}$ & $3.917\times10^{-4}$ & $5.567\times10^{-6}$ \\
 & 0.5 & $1.238\times10^{42}$ & $6.528\times10^{-4}$ & $9.278\times10^{-6}$ \\
\hline
{1.6 $M_\odot$} & 0.1 & $2.475\times10^{41}$ & $1.157\times10^{-4}$ & $1.645\times10^{-6}$ \\
 & 0.3 & $7.426\times10^{41}$ & $3.471\times10^{-4}$ & $4.934\times10^{-6}$ \\
 & 0.5 & $1.238\times10^{42}$ & $5.786\times10^{-4}$ & $8.223\times10^{-6}$ \\
\hline
{1.8 $M_\odot$} & 0.1 & $2.475\times10^{41}$ & $1.016\times10^{-4}$ & $1.444\times10^{-6}$ \\
 & 0.3 & $7.426\times10^{41}$ & $3.048\times10^{-4}$ & $4.331\times10^{-6}$ \\
 & 0.5 & $1.238\times10^{42}$ & $5.079\times10^{-4}$ & $7.219\times10^{-6}$ \\
\hline
{2.0 $M_\odot$}  & 0.1 & $2.475\times10^{41}$ & $9.136\times10^{-5}$ & $1.298\times10^{-6}$ \\
 & 0.3 & $7.426\times10^{41}$ & $2.741\times10^{-4}$ & $3.895\times10^{-6}$ \\
 & 0.5 & $1.238\times10^{42}$ & $4.568\times10^{-4}$ & $6.492\times10^{-6}$ \\
\hline
{2.2 $M_\odot$} & 0.1 & $2.475\times10^{41}$ & $8.780\times10^{-5}$ & $1.248\times10^{-6}$ \\
 & 0.3 & $7.426\times10^{41}$ & $2.634\times10^{-4}$ & $3.744\times10^{-6}$ \\
 & 0.5 & $1.238\times10^{42}$ & $4.390\times10^{-4}$ & $6.239\times10^{-6}$ \\
\hline
\multicolumn{5}{c}{\((\Delta \Omega_{gl} / \Omega)_{\text{obs}} = 1.38 \times 10^{-6}\)} \\
\hline\hline
\end{tabular}
\caption{The angular momentum transferred $\Delta \mathrm{L}$, the change in the CS rotational speed $\Delta \Omega_{gl}$, and corresponding glitch amplitude $\Delta \Omega_{gl} / \Omega$  are calculated for different masses and for different values of the pairing gap.}
\label{tab:glitch_mass_diff}
\end{table}
Using the constructed EOS, we calculate the superfluid angular momentum transferred ($\Delta L$) and the resulting change in angular velocity ($\Omega_{\text{gl}}$). The glitch amplitude, defined as $\Delta \Omega_{gl}/\Omega$, is computed using the Vela pulsar’s spin frequency, $\Omega \approx 70$ Hz. These results are summarized in Table \ref{tab:glitch_mass_diff}. 
Initial theoretical estimations predicted that neutron pairing inside the core of the superfluid is in $\prescript{3}{}{P}_2$ state \cite{Hoffberg1970,Takatsuka_1971,Takatsuka_1972}. The neutron superfluid pairing gaps in the stellar core are adopted from \cite{Sedrakian:2018ydt,Ramanan:2020xrj}, and the final row of the table provides a representative mean glitch amplitude derived from observational data across multiple pulsars, as reported in \cite{arumugam2023classification}. 

Table \ref{tab:glitch_mass_diff} presents the angular momentum transfer and the resulting glitch amplitudes for hybrid stars with masses ranging from $1.4\,M_\odot$ to $2.2\,M_\odot$ and for different values of the pairing gap. For a fixed stellar mass, an increase in the pairing gap enhances the angular momentum reservoir, leading to larger changes in the rotational frequency and hence larger glitch amplitudes. For a given pairing gap, the glitch amplitude decreases with increasing stellar mass, since more massive stars possess a larger total moment of inertia. As a result, the same amount of angular momentum transfer produces a smaller fractional spin-up. Nevertheless, the variation in $\Delta\Omega_{\rm gl}/\Omega$ is modest and remains of the order $\sim10^{-6}$, indicating that the glitch mechanism is relatively insensitive to the global stellar mass and is mainly governed by the microphysics of the mixed phase.

The correlation between the pinning force per unit length and the surface tension plays an important role in determining the glitch amplitude. In the present work, we have considered a simplified proportional relation between these two quantities to capture the overall dependence of the glitch magnitude on the microphysical properties of the mixed phase. Table \ref{tab:fp_amplitude} shows the variation of $\Delta\Omega_{\rm gl}/\Omega$ with pinning force $f_p$ for different neutron superfluid gaps. It is observed larger $f_p$ values produce higher glitch magnitudes, while those with smaller $f_p$ yield weaker glitches. Our calculated values show a close agreement with these typical giant glitch amplitudes.

\begin{table}[htbp]
\centering
\renewcommand{\arraystretch}{1.2}
\begin{tabular}{c c c c}
\hline\hline
$f_p$
& $\Delta\Omega_{\rm gl}/\Omega$ 
& $\Delta\Omega_{\rm gl}/\Omega$ 
& $\Delta\Omega_{\rm gl}/\Omega$  \\
&($\Delta=0.1$ MeV)&($\Delta=0.3$ MeV)&($\Delta=0.5$ MeV)\\
\hline
$1.0\,\sigma$ & $2.44\times10^{-6}$ & $7.31\times10^{-6}$ & $1.22\times10^{-5}$ \\
$0.9\,\sigma$ & $2.19\times10^{-6}$ & $6.58\times10^{-6}$ & $1.10\times10^{-5}$ \\
$0.8\,\sigma$ & $1.95\times10^{-6}$ & $5.85\times10^{-6}$ & $9.75\times10^{-6}$ \\
$0.7\,\sigma$ & $1.71\times10^{-6}$ & $5.12\times10^{-6}$ & $8.53\times10^{-6}$ \\
$0.6\,\sigma$ & $1.46\times10^{-6}$ & $4.39\times10^{-6}$ & $7.31\times10^{-6}$ \\
$0.5\,\sigma$ & $1.22\times10^{-6}$ & $3.66\times10^{-6}$ & $6.09\times10^{-6}$ \\
$0.4\,\sigma$ & $9.75\times10^{-7}$ & $2.92\times10^{-6}$ & $4.87\times10^{-6}$ \\
$0.3\,\sigma$ & $7.31\times10^{-7}$ & $2.19\times10^{-6}$ & $3.66\times10^{-6}$ \\
$0.2\,\sigma$ & $4.87\times10^{-7}$ & $1.46\times10^{-6}$ & $2.44\times10^{-6}$ \\
$0.1\,\sigma$ & $2.44\times10^{-7}$ & $7.31\times10^{-7}$ & $1.22\times10^{-6}$ \\
\hline\hline
\end{tabular}
\caption{The dependence of the glitch amplitude $\Delta\Omega_{\rm gl}/\Omega$ on the pinning force $f_p$ and correlation of surface tension $\sigma$ is shown for different neutron superfluid gaps in a 1.44 $M_{\odot}$ HS.}
\label{tab:fp_amplitude}
\end{table}

We get the time interval between consecutive glitches in the Vela pulsar as $\sim 2.5$ years, closely matching the observed time interval, which is $\sim 3$ years.

\section{Summary and conclusion} \label{sec:conclusion}
Glitches are sudden spin-ups in the rotational frequencies of rotating neutron stars. Most of the glitches are of the order of $\Delta \Omega/\Omega \sim 10^{–9}$. Some of the glitches are large. Most of the large glitches detected are from the Vela pulsar falling range of $\Delta \Omega/\Omega \sim 10^{–6}$. Moreover, these substantial glitches disrupt the rotation of the Vela pulsar approximately every three years, a notably high frequency compared to most pulsars, which typically experience fewer than one such event per decade \cite{Espinoza_2011,Espinoza_2021}.

The seminal works of Anderson and Itoh claimed that glitches are manifestations of neutron superfluidity inside neutron stars, which is a high-density fermionic system. This explanation is based on the pinning of superfluid vortices in the crust of the star.

However, this reasoning encounters challenges when accounting for the nondissipative entrainment coupling between the neutron superfluid and the crustal lattice, an interaction that can be described by an effective neutron mass, $m^*_n$ \cite{Chamel_2012}. Studies suggest that this effective mass may be considerably larger than the bare neutron mass, $m_n$. This increased effective mass reduces the superfluid's mobility relative to the lattice, necessitating a larger angular momentum reservoir to produce observable glitches \cite{Andersson_2012}. 

In view of this, we attempt to include core superfluidity to account for large glitches. We consider the star to be an HS containing pure NM in the outer core region and a mixed phase of NM and deconfined SQM in the inner core.  
The mixed phase region consists of a lattice structure between the nucleonic phase and the quark phase. The surface tension associated with these structures is responsible for vortex pinning. 
The large surface tension associated with quark structures provides a correspondingly strong vortex pinning force, allowing a substantial angular momentum reservoir to be stored in the pinned superfluid.
With the decrease of star rotational speed, the Magnus force unpins the pinned vortices, transferring the stored angular momentum in the vortices to the star. The release of this accumulated angular momentum during vortex unpinning events naturally leads to large glitch amplitudes. This highlights the crucial role of quark–hadron interface physics in explaining giant glitches in hybrid stars.

As the present work primarily focuses on the magnitude of the spin-up 
(i.e., glitch amplitude) in the star's rotational frequency, a natural 
extension of this study would be to model the subsequent postglitch 
spin-down evolution. In the framework proposed here, the spin-down 
relaxation is expected to occur due to the gradual repinning and 
recoupling of neutron superfluid vortices to the quark structures in 
the mixed phase region. 
Such a process  may lead to 
a measurable change in the post-glitch spin-down rate $\dot{\Omega}$. A quantitative treatment of this phenomenon requires solving the 
time-dependent rotational evolution equations which will be addressed in future investigations.

Another aspect worth discussing concerns high-mass compact stars, 
for which the central density may exceed the threshold for the 
formation of a pure quark matter core. Depending on the pairing 
pattern, such a core may exist in normal quark matter or in a 
color-superconducting phase such as the two-flavor superconducting or color–flavor locked phase \cite{Barrois1977,Alford2008,Iida2001}.
If the pure quark core is in a superfluid state, it can sustain 
quantized rotational vortices and therefore act as an additional 
angular momentum reservoir \cite{Iida2002}. This introduces the possibility of 
multiple dynamically distinct superfluid components within the star; 
one associated with neutron superfluidity in the mixed 
phase region, and another associated with quark superfluidity in the 
core.
The presence of a quark superfluid reservoir could have several 
potential implications. First, it may increase the total angular 
momentum available for transfer during a glitch, possibly leading 
to larger glitch amplitudes in sufficiently massive stars. Second, 
differences in pinning strength and coupling timescales between the 
quark core and the surrounding matter could give rise to multistage 
angular momentum transfer or complex postglitch relaxation behavior. 
Finally, the interaction between neutron vortices and quark vortices 
at the interface may introduce additional dynamical effects, which 
could further enrich the glitch phenomenology.

We also note that the present hadronic model does not include heavy baryonic 
degrees of freedom such as $\Delta$ resonances, which may appear at 
densities of order $\sim 2 n_0$ in the neutron star \cite{Oliveira:2007zz,Drago2014,JiaJie2018,Vishal2025}. The inclusion of such species would generally soften the 
equation of state and modify the stellar mass–radius relation, 
potentially affecting the size of the mixed phase region \cite{Malfatti2020}.
Nevertheless, in the present framework the dominant contribution to 
the pinning force originates from the surface tension of the quark 
structures. As seen from Eq. (\ref{eq:surface tension}), the surface tension depends 
primarily on the Fermi momenta of the quarks. Since the glitch 
amplitude scales with the pinning force, it is largely governed by 
mixed-phase microphysics. Therefore, although the appearance of 
$\Delta$ resonances may introduce quantitative corrections through 
structural changes, we do not expect substantial qualitative 
modifications of the predicted glitch amplitudes.

\section*{Acknowledgments}

The authors gratefully acknowledge the financial support provided by the Science and Engineering Research Board (SERB), Department of Science and Technology, Government of India, under Project No.\ CRG/2022/000069. The authors sincerely thank Professor Debades Bandyopadhyay for his valuable discussions and insightful suggestions. They also acknowledge Xia ChengJun for fruitful discussions. Finally, the authors express their gratitude to the anonymous referee for the constructive comments and suggestions, which have significantly improved the quality of this manuscript.

\bibliography{references}
\end{document}